\documentclass[conference,a4paper]{IEEEtran}
\IEEEoverridecommandlockouts
\usepackage{cite}
\usepackage{amsmath,amssymb,amsfonts}
\usepackage{algorithmic}
\usepackage{graphicx}
\usepackage{textcomp}
\usepackage{xcolor}
\usepackage{hyperref}
\usepackage{url}
\usepackage{soul} % for the command 'hl'
\usepackage{xspace}
\usepackage{multirow}
\usepackage{balance}
\usepackage{fancyhdr} 

\def\VITAMINV{\mbox{Vitamin-V}\xspace}
\def\RISCV{\mbox{RISC-V}\xspace}
\def\BibTeX{{\rm B\kern-.05em{\sc i\kern-.025em b}\kern-.08em
    T\kern-.1667em\lower.7ex\hbox{E}\kern-.125emX}}

\fancypagestyle{firstpage}{%
 \fancyhf{}%
 \fancyhead[C]{\textit{Special Session: Vitamin-V Workshop}}
}

\begin{document}

\title{Security and RAS in the Computing Continuum\thanks{Funded by the European Union under the Horizon Europe Programme. Project name: \VITAMINV. Project number: 101093062. Views and opinions expressed are, however, those of the authors only and do not necessarily reflect those of the European Union or the HaDEA. Neither the European Union nor the granting authority can be held responsible for them. This paper is also funded by HFRI with title REDESIGN and Project Number 16973 and project SERICS (PE00000014) under the MUR National Recovery and Resilience Plan funded by the European Union - NextGenerationEU.}}

\author{\IEEEauthorblockN{Martí	Alonso\textsuperscript{1}, David Andreu\textsuperscript{1}, Ramon Canal\textsuperscript{1}, Stefano Di Carlo\textsuperscript{2}, Odysseas Chatzopoulos\textsuperscript{3},\\Cristiano Chenet\textsuperscript{2}, Juanjo Costa\textsuperscript{1}, Andreu Girones\textsuperscript{1}, Dimitris Gizopoulos\textsuperscript{3}, George Papadimitriou\textsuperscript{3},  \\ Enric Morancho\textsuperscript{1}, Beatriz Otero\textsuperscript{1}, 
Alessandro Savino\textsuperscript{2}}
\IEEEauthorblockA{\\
\textsuperscript{\textbf{1}}Universitat Politècnica de Catalunya, Barcelona, Spain \\
\textsuperscript{\textbf{2}}Politecnico di Torino, Torino, Italy \\
\textsuperscript{\textbf{3}}University of Athens, Athens, Greece \\ 
Contact email: ramon.canal@upc.edu}}

\maketitle
\thispagestyle{firstpage}  % Apply fancy header on the first page

\begin{abstract}
Security and RAS are two non-functional requirements under focus for current systems developed for the computing continuum. Due to the increased number of interconnected computer systems across the continuum, security becomes especially pervasive at all levels, from the smallest edge device to the high-performance cloud at the other end. Similarly, RAS (Reliability, Availability, and Serviceability) ensures the robustness of a system towards hardware defects. Namely, making them reliable, with high availability and design for easy service. 

In this paper and as a result of the \VITAMINV EU project, the authors detail the comprehensive approach to malware and hardware attack detection; as well as, the RAS features envisioned for future systems across the computing continuum.
\end{abstract}


%vkarakos: we need to be consistent in the use of Vitamin-V 

\bstctlcite{IEEEexample:BSTcontrol}

\section{Introduction}

\RISCV is a revolutionary open-source instruction set architecture (ISA) designed to offer simplicity, modularity, and extensibility \cite{Waterman:EECS-2016-129}. This exciting development brings many benefits over proprietary processor architectures, including the potential for customization and lower licensing costs \cite{greengard2020will}.

Despite these advantages and the fact that \RISCV applications have started to see their birth in the embedded domain \cite{10.1145/3457388.3458657}, several challenges still need to be addressed before \RISCV can be widely adopted beyond conventional ones like performance or standardization.
Among these challenges, security and RAS are of utmost importance. As \RISCV processors become more widely adopted, there is an increasing potential for security attacks. Ensuring the security of \RISCV-based applications will be an important challenge that needs to be addressed as technology develops.

Another key challenge is RAS. Enterprise and cloud data centers are increasingly integrating complex System-on-Chip (SoC) architectures to meet the demands of modern computing workloads. However, the widespread deployment of these devices raises the risk of undetected faults, which can lead to critical issues like system crashes or silent data corruptions (SDCs). These faults, stemming from manufacturing defects and in-field reliability concerns, present significant challenges for data centers. While cosmic ray-induced soft errors have been extensively studied \cite{9894291,10071105,9499847,9042036,9408178,9857727,10151692,9616430}, modern data centers must also consider other potential fault sources, such as manufacturing defects and thermal variations \cite{9296366}, which may lead to SDCs during normal operations \cite{9983919,sdcs-fb,10.1145/3458336.3465297,10.1145/3600006.3613149,gizopoulos_abc}.

%In this context, 
This paper presents AI-based malware and hardware attack detection. With special emphasis on the reproducible and cross-platform methodology; as well as RAS.

The paper is organized as follows: Section \ref{sec:MethodologyAI} describes the step-by-step methodology used for AI-based security and Section \ref{sec:MethodologyRAS} provides it for RAS. Then, Section \ref{sec:Results} presents the performance and detection capabilities of the AI-based security and the RAS analysis. Finally, Section \ref{sec:conclusions} summarizes the main contributions of the paper.

\section{Methodology for AI-based security}
\label{sec:MethodologyAI}

This section describes the methodology for the AI-based malware and hardware attack detection. We propose two different sources of information to detect the malign behaviour: hardware performance monitors (HPM) and instruction opcodes. While hardware performance monitors are only available during the execution of the program (thus, dynamically), opcodes are both available statically and dynamically. Consequently, in this section we describe the process of using each of these sources of information to process through ML algorithms to detect the attacks.

\subsection{Hardware Performance Monitoring }
%UOA

Modern processors include hardware performance monitoring units to track the CPU performance, necessary due to increased processor complexity, such as hierarchical cache subsystems, non-uniform memory, and out-of-order execution. Software that adapts to resource utilization benefits from better performance and efficiency.
The HPM includes registers and counters for microarchitectural events accessible by the Operating System (OS) using libraries like Linux perf \cite{rperf} and PAPI \cite{1592755}.  

HPMs track events such as retired instructions, branch predictions, cache hits/misses, and clock ticks, but only a limited number of hardware counters can be active at a time due to design and cost constraints \cite{Sprunt_2002,Malone_2011,Doyle_2017}. Developers access counters via performance monitoring instructions, reading, and writing counter values.

\subsection{Static Analysis of applications}
%UPC 

%\begin{figure}[bt]
%    \centering
%  \includegraphics[width=.75\linewidth]{figures/staticPhases.pdf}
%    \caption{Methodology used to create the Deep Learning model.}
%  \label{fig:staticPhases}
%\end{figure}

Evaluating software trustworthiness early is essential. We developed a machine learning (ML) tool that analyzes static executable content to determine if it is benign or malicious. Inspired by previous work on detecting software bugs and security threats using static analysis \cite{ding2014control, haddadpajouh2018deep}, we incorporated deep learning (DL) techniques to identify complex patterns. In cases where the dataset is insufficient or a zero-day attack is involved, transfer learning (TL) can be used to enhance detection accuracy by leveraging pre-trained models \cite{Rodriguez2022}.

Following Haddadpajouh et al. \cite{haddadpajouh2018deep} approach, we breakdown the process in four steps: dataset creation, feature extraction, model training, and deployment. To train the model, we gather a balanced dataset of benign and malicious programs, including hardware attacks like Spectre \cite{Kocher2019}, Meltdown \cite{Lipp2018}, viruses, and malware. Malicious programs will be sourced from platforms like VirusTotal \cite{VirusTotal}, VirusShare \cite{VirusShare}, or SourceFinder \cite{rokon2020sourcefinder}. Benign programs will come from regular Linux applications in the Debian repository.

Programs are converted to feature vectors by disassembling ELF binaries to extract operation codes (OpCode) sequences. These sequences are analyzed to generate feature vectors to train the DL model. The model developed and benchmarked in Section \ref{sec:Results} uses AMD64. Yet, the same methodology can be applied to other ISAs.

\subsection{Dynamic Analysis using Hardware Performance Counters}

%\begin{figure}[htb]
%\centerline{\includegraphics[width=\columnwidth]{figures/Anomaly_detector_framework}}
%\caption{The hardware and AI-based anomaly detector framework.}
%\label{fig:Anomaly_detector_framework}
%\end{figure}

Anomaly detection using hardware monitoring and AI involves dynamic analysis of microarchitectural events via ML algorithms to identify abnormal behavior. This approach, first introduced by Demme et al. in 2013 \cite{Demme_2013}, has been applied to malware detection \cite{Chenet_2023} and hard and soft errors \cite{9474120,kasap2023microarchitectural} but not on \RISCV.

Programs exhibit phase behaviors \cite{Sherwood_2003,Isci_2006}, allowing for anomaly detection through patterns in hardware performance counters. The proposed anomaly detection framework includes three main components: (i) a CPU with HPM, (ii) data collection, and (iii) anomaly detection via ML classifiers. Challenges arise due to the cost and complexity of monitoring speculative execution events, especially in resource-constrained devices, where balancing hardware events and detection accuracy is critical. To overcome limited counter availability, some methods run applications multiple times to capture more events \cite{Demme_2013,Singh_2017,Sayadi_2017}, but this impacts runtime applicability.

Data collection involves selecting events, extracting features, and reducing dimensions \cite{Chenet_2023}. Feature extraction captures HPCs into vector space, while dimensionality reduction minimizes redundant data that can decrease detection accuracy. When empirical event selection is not feasible, techniques like Principal Component Analysis (PCA), Fisher Score, and Information Gain are used to identify relevant features \cite{Peng_2005}. The final block, anomaly detection, is carried out by ML classifiers. These classifiers can vary by type (multi-class or one-class), learning method (supervised, unsupervised, semi-supervised), or underlying algorithm (Neural Networks, Decision Trees, etc.). Multi-class classifiers assume multiple normal classes, while one-class classifiers identify anomalies based on a single normal class boundary \cite{Chandola_2009}. Given the challenges of labeling data, unsupervised learning is often preferred \cite{Chandola_2009}. To improve accuracy, advanced methods like Ensemble Learning \cite{Sayadi_2018}, Boosting \cite{Freund_Schapire_1997}, and multi-stage classifiers \cite{Sayadi_2019} are used.

\section{Methodology for RAS}
\label{sec:MethodologyRAS}

\subsection{Reliability, Availability, and Serviceability in Large-Scale SoC Deployment}

Detecting and managing faults that result in SDCs is particularly challenging due to the specific conditions required for them to manifest \cite{itc_disneyland,10616056,10567770}. These conditions can include particular machine instruction sequences, variations in operating voltages \cite{papadimitriou2021systemlevel}, temperature fluctuations, and platform-level behaviors like interrupts. This complexity results in low repeatability in SDC testing, necessitating extended testing periods to uncover potential issues. Developing effective testing methodologies to identify and address SDCs is therefore crucial for maintaining reliability, availability, and serviceability (RAS) in data centers. These strategies may involve repeated execution of specific code sequences to trigger SDCs or the use of pseudo-random instruction sequences to increase variability and expose latent faults during testing \cite{10609719}.

\subsection{Sources of Faults and Impact on Reliability}

SoCs can experience faults from various sources, including radiation, electrical marginalities, and silicon defects that may not be detectable during manufacturing. These issues may manifest in the field, impacting system reliability. The effects of such faults vary depending on where they occur within the SoC circuitry. For example, faults in error detection and correction-equipped circuits, such as caches, can be corrected at the hardware level, preventing system disruption \cite{10.1145/3613424.3614304}. However, SDCs where data errors propagate without triggering an interruption can have unpredictable consequences depending on the application \cite{10224870}. A minor data error in a graphical operation might be negligible, but the same issue in a financial transaction could have serious implications.

Managing SDCs at scale is particularly crucial when deploying millions of processing cores, as even a single defect can cause significant operational disruptions. For instance, a modest-sized data center with 100,000 SoCs could experience at least one SDC per month at a 10 failures in time (FIT) rate \cite{9983919} (where 1 FIT corresponds to one failure per billion hours of operation) \cite{10567770,10609719}. This occurrence rate underscores the need for rigorous reliability, availability, and serviceability (RAS) strategies to mitigate the effects of SDCs in large-scale computing environments. Hyperscale data centers, with millions of deployed SoCs, face a more acute risk, making reliability assessments and fault management an ongoing priority.

\subsection{RAS-Oriented Approach}

Ensuring the reliability, availability, and serviceability of data center infrastructure requires more than simply detecting data corruption and defective chips. It demands a comprehensive strategy that encompasses both hardware and software design. Reliability can be enhanced by identifying and isolating defective components through regular testing, while availability ensures systems continue to function smoothly even in the presence of faults. Serviceability emphasizes quick fault diagnosis and repair to minimize downtime and provide useful information for system debug and defect root causing.

One effective strategy for early-stage mitigation is simulation-based testing. Researchers can use architectural and microarchitectural models to simulate SoC behavior under various conditions, identifying potential faults before physical chips are manufactured. These models allow prediction of FIT rates early in the design process, enabling adjustments when they are less costly. However, early-stage models only approximate final hardware behavior, as they often lack complete design details.

In the late stages of development, more precise measurements can be made using gate-level models and actual silicon. However, these methods are more resource-intensive and time-consuming, limiting their applicability to scenarios where extremely high accuracy is required, such as in critical systems or hyperscale environments. By combining early predictive models with late-stage testing, data centers can implement a robust RAS framework to proactively address SDC risks.

\section{Results}
\label{sec:Results}

\subsection{AI-based security}
The preliminary results on detecting malware and hardware attacks using HPCs are presented in this section, leveraging both supervised and unsupervised learning classifiers.
\subsubsection{Supervised learning}
With supervised classifiers, the detection of a specific type of hardware attack, i.e., side-channel attacks, is analyzed during runtime by monitoring HPCs.

\subsubsection*{Dataset creation}
We selected 7 hardware-based attacks (i.e., Meltdown~\cite{Lipp2018}, Spectre~\cite{Kocher2019}(V1, V2 and V4),  ZombieLoad~\cite{Schwarz2019ZombieLoad},   Fallout~\cite{canella2019fallout} and Crosstalk~\cite{ragab_crosstalk_2021}) and 7 benign programs (i.e., Matrix multiplier, Debian stress tool~\cite{stress_git}, MiBench
Bitcount~\cite{mibench}, STREAM benchmark~\cite{stream}, bzip2~\cite{bzip2} and FFmpeg~\cite{ffmpeg}) to construct the dataset.

We executed these applications and recorded multiple HPCs during binary execution using the \textit{perf} tool. To streamline operations, we restricted execution to a single CPU core
using the \verb|taskset| Linux tool, ensuring that the collected HPC data remains unaffected by workload distribution across cores.

Some hardware attacks, like those in the Spectre family, exploit wrong speculative execution, triggered when the branch predictor mispredicts the outcome of a branch instruction. Therefore we selected both \verb|branch-instructions| and \verb|branch-misses| generic \textit{perf} events, which provide the ratio between total branches and those where the predictor missed. This selection is supported by previous work, such as Congmiago Li et al.~\cite{9437667}.

Additionally, side-channel hardware-based attacks heavily stress the computer's cache memory. A high count of cache misses on the last-level cache (LLC) memory may indicate the presence of a FLUSH+RELOAD side-channel attack, known as the most effective and popular among hardware-based attacks. The first-level cache is also a common target in other
attacks, as used by Stefano Carnà et al.~\cite{10.1145/3519601}. Thus, we also selected the \verb|LLC-load-misses| and \verb|L1-dcache-load-misses| events.

Previous works have used sampling rates ranging from 1 ms per sample to 100 ms per sample. Congmiago Li et al.~\cite{9437667} even dynamically change the sample rate to prevent evasive malware. To generate a large number of samples for the machine learning model, the aim was to use the lowest possible sample rate. However, experimental results showed that rates below 10 ms caused anomalies in \textit{perf}, such as missed samples. As a result, we chose a 1 ms sample rate. We extract 2,000 samples from each application, either by running the application for 2 seconds or repeating the execution until we get those samples.

\subsubsection*{Dataset preparation}

The dataset for the 14 applications contains 28,000 samples, we created 3 different scenarios: Balanced, Only malign, and Only benign. Each scenario was split using 80\% for training and 20\% for testing (as shown in Table~\ref{tab:dataset_size}).

\begin{table}[t]
	\caption{Samples distribution for the 3 scenarios: Balanced, Only malign, and Only benign.}
	\label{tab:dataset_size}
  \centering
    \begin{tabular}{l rr rr}
        \multirow{2}{*}{Dataset} & \multicolumn{2}{c }{Train}           & \multicolumn{2}{c}{Test}            \\ \cline{2-5}
                                & \multicolumn{1}{c }{Malign} & Benign & \multicolumn{1}{c }{Malign} & Benign \\ \hline
        \textbf{Balanced}       & \multicolumn{1}{r }{11200}  & 11200  & \multicolumn{1}{r }{2800}   & 2800   \\
        \textbf{Malign}         & \multicolumn{1}{r }{11200}  & 0      & \multicolumn{1}{r }{2800}   & 14000  \\
        \textbf{Benign}         & \multicolumn{1}{r }{0}      & 11200  & \multicolumn{1}{r }{14000}  & 2800   \\ \hline
    \end{tabular}
\end{table}

\subsubsection*{Model evaluation}
%include graphics results for SVM

%include graphics results for RF
\begin{figure*}[h]
    \centering
    \parbox{0.3\textwidth}{
      \includegraphics[width=\linewidth]{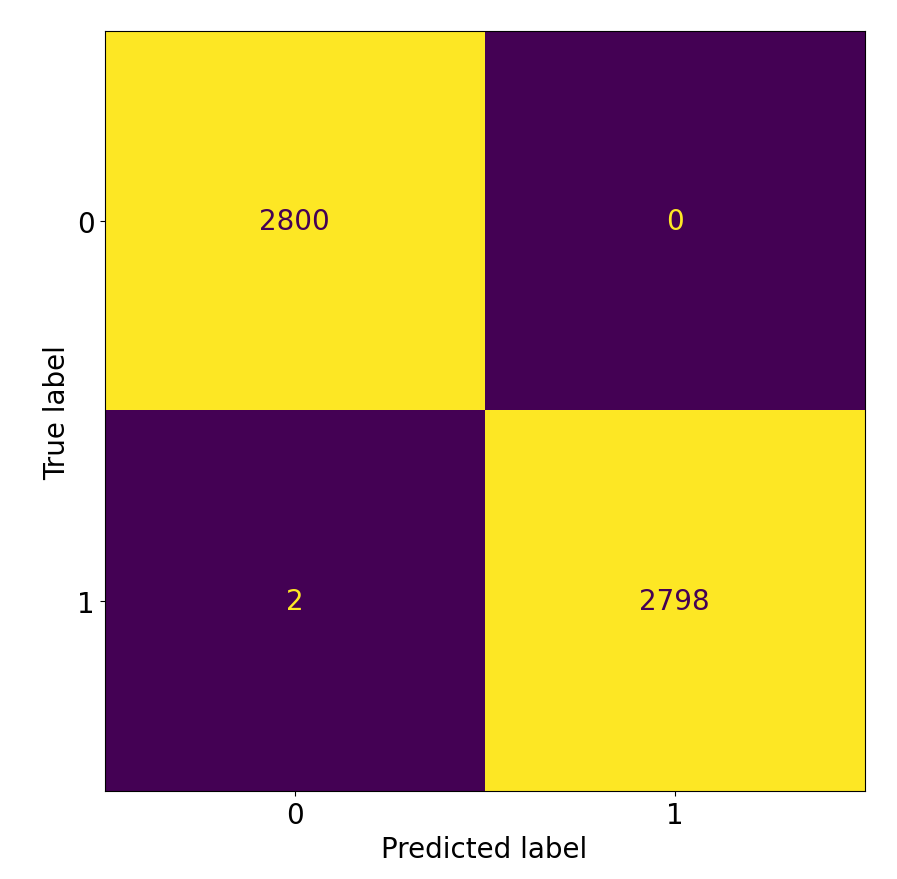}
    }
    \qquad
    \parbox{0.3\textwidth}{
      \includegraphics[width=\linewidth]{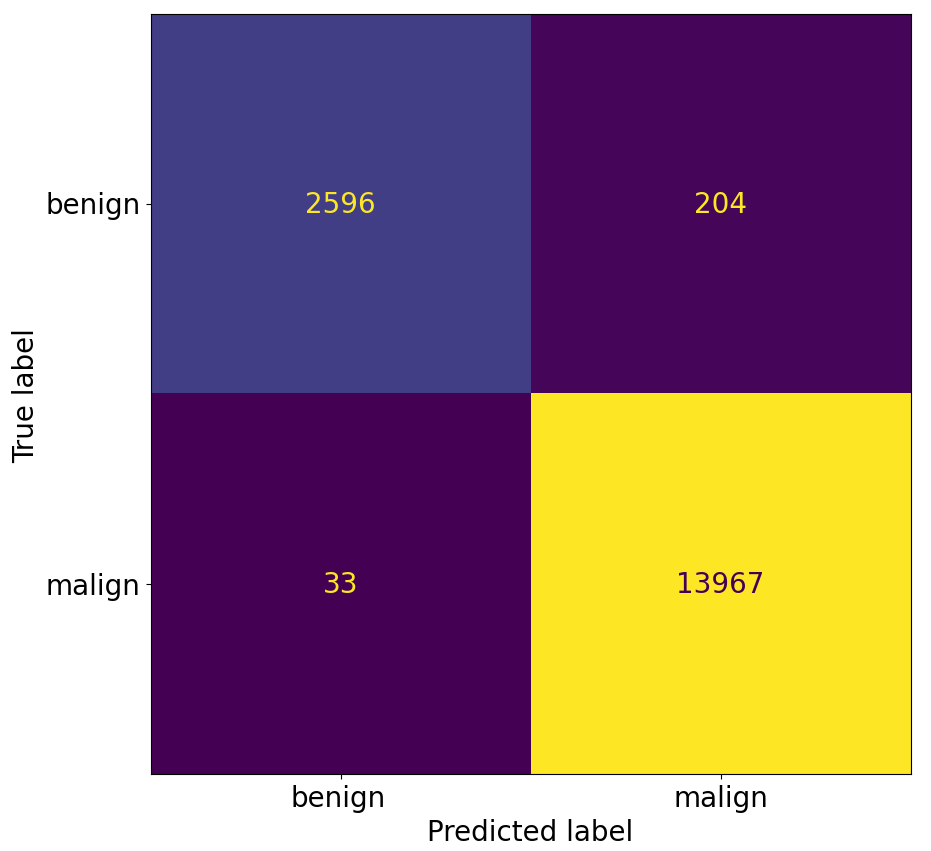}
    }
    \qquad
    \parbox{0.3\textwidth}{
      \includegraphics[width=\linewidth]{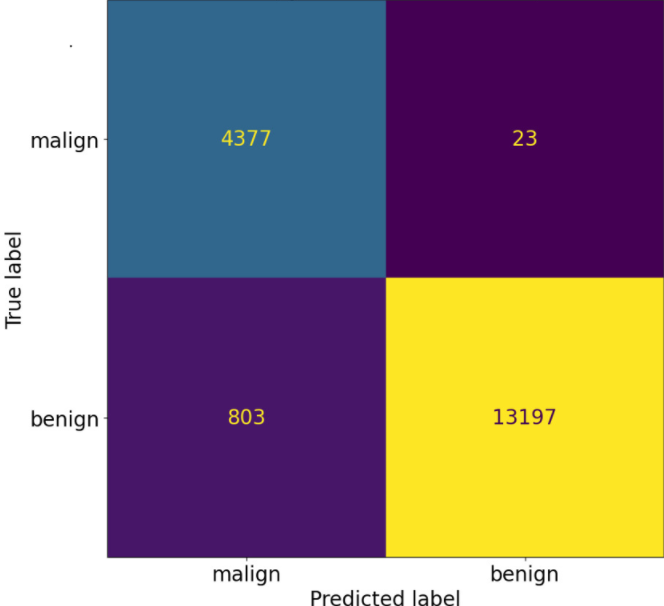}
    }
    \caption{Confusion matrices for detecting malware using a balanced data set (left),
    a benign dataset (center) and a malign dataset (right).}
    \label{fig:detect_OCSVM}
\end{figure*}

%include graphics results for RF
\begin{figure*}[t]
    \centering
     \includegraphics[width=0.75\linewidth]{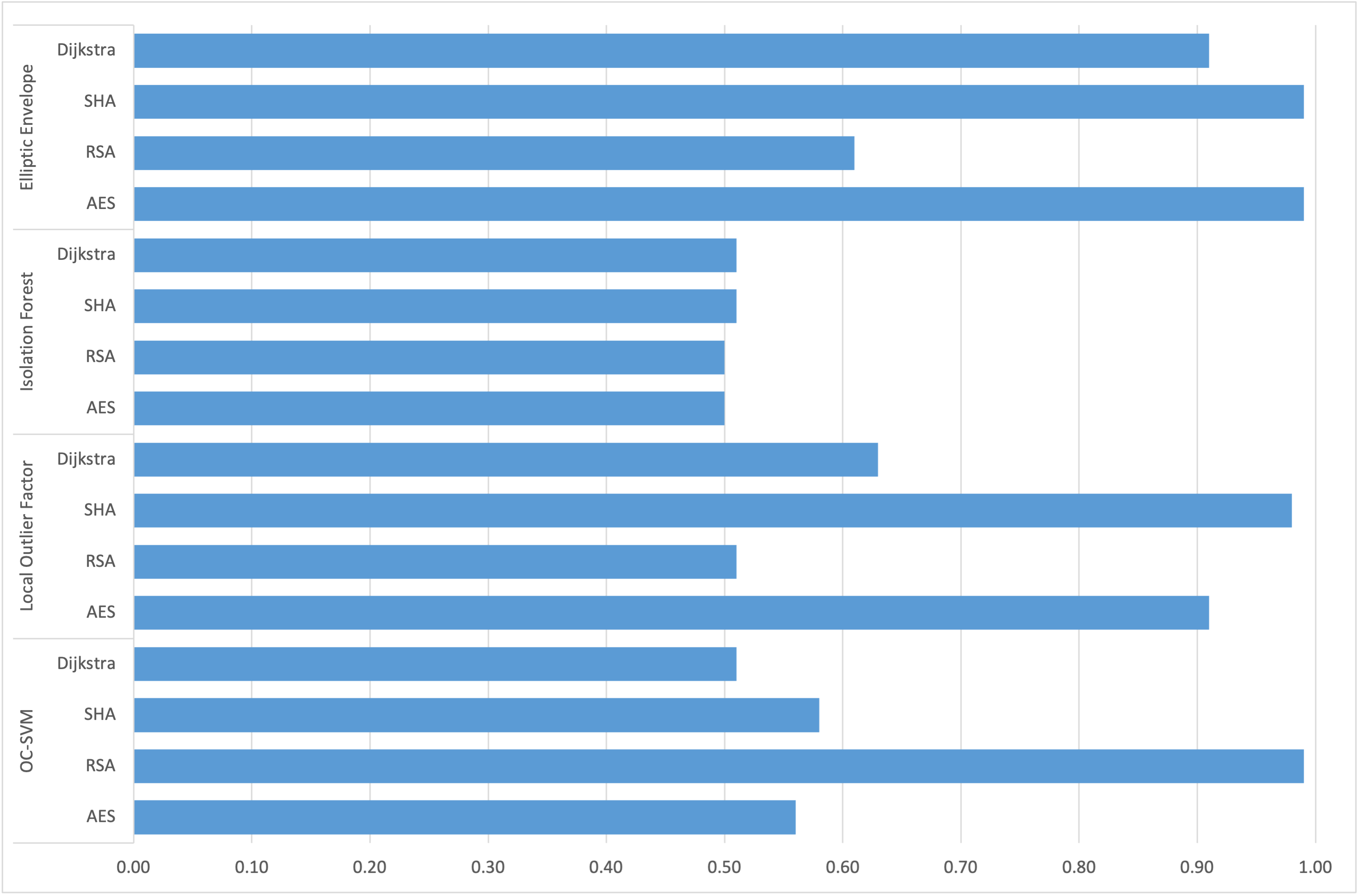}
    \caption{Accuracy of unsupervised SBO detection for different benchmarks and classifiers. The malicious function runs a number of instructions lower than 1\% of those of the original application.).}
    \label{fig:unsupervised}
\end{figure*}

We used different models for training: Support Vector Machine~(SVM) and One-Class SVM. The SVM model shows the best results for detecting side-channel attacks with a 99\% accuracy, where only 2 of the attacks were misdetected (as observed in the confusion matrix of Figure~\ref{fig:detect_OCSVM} left).

%\begin{figure*}[t]
%    \centering
%    \includegraphics[width=0.25\textwidth]{figures/detection_SVM_RBF.png}
%    \caption{Confusion matrix for detection using a balanced dataset.}
%    \label{fig:detect_SVM}
%\end{figure*}

Figure~\ref{fig:detect_OCSVM} (center and right) shows the scenarios where a single class dataset is used for training (either benign or malign). In these cases, we observed many false positives
because the model encounters samples that were unseen during training and therefore misclassifies them as malware.

\subsubsection{Unsupervised learning}

To evaluate the effectiveness of unsupervised learning techniques, an experiment was designed to detect Stack Buffer Overflow (SBO) attacks. Four applications from the MiBench suite were selected—AES, RSA, SHA, and Dijkstra—and were subjected to SBO attacks. These attacks triggered the execution of a malicious function, with its size parametrized relative to the original code (i.e., the smaller the malicious function, the more stealthy its execution).

For each application, 20,000 executions were collected (10,000 for training and 10,000 for testing), with benchmark inputs randomly varied. The training dataset consisted solely of benign executions, while the testing dataset included 50\% benign executions and 50\% with attacks. Four anomaly detection models—OC-SVM, LOF, Isolation Forest, and Elliptic Envelope—were trained to assess their effectiveness on the dataset. Preliminary results, shown in Figure~\ref{fig:unsupervised}, indicate promising performance for the Elliptic Envelope classifier, except the RSA benchmark, which remains under investigation.

\subsection{RAS}
This subsection showcases the effects of \textit{Permanent} faults on program execution. We focus on SDC outcomes, showing the probability that a fault in a specific hardware unit results in an SDC. The basic parameters of the gem5 configuration we use in this paper can be seen in Table~\ref{tab:basic_params}.

\begin{scriptsize}
\begin{table}[t]
    \renewcommand{\arraystretch}{1.2}
    \vspace{-4mm}
    \centering
     \caption{Major Simulator Configurations for each ISA.} 
    \label{tab:basic_params}
    \vspace*{-1mm}
    % Check --> Excel2latex 
     \footnotesize{
        \begin{tabular}{|c|c|}
            \hline
            \textbf{Parameter} & \textbf{Value} \\
            \hline
            \hline
            ISA & RISC-V / Arm / x86 \\
            \hline
            Pipeline & 64-bit OoO (8-issue) \\
            \hline
            L1 Instruction Cache & 32KB, 64B line, 128 sets, 4-way \\ 
            \hline
            L1 Data Cache & 32KB, 64B line, 128 sets, 4-way\\ 
            \hline
            L2 Cache & 1MB, 64B line, 2048 sets, 8-way \\ 
            \hline
            Physical Register File & 128 Int; 128 FP \\
            \hline
            LQ/SQ/IQ/ROB entries & 32/32/64/128  \\
            \hline
        \end{tabular}
    }   
\end{table}
\end{scriptsize}

\subsubsection{SDCs due to Permanent Faults in L1 instruction and data caches}

%\subsubsection{L1 Instruction Cache}

\begin{figure}[t]
    \centering
    %\vspace*{-1mm}
    \includegraphics[width=0.5\textwidth]{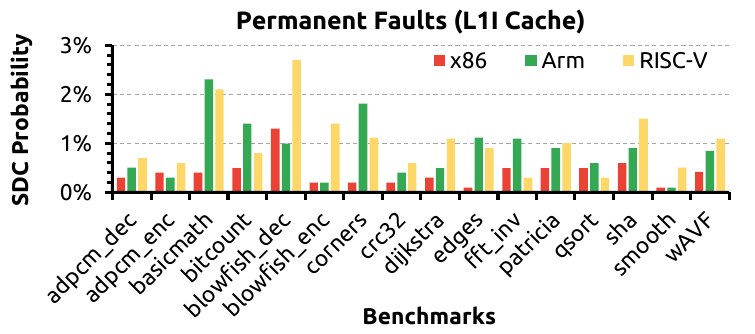}
    \vspace*{-4mm}
    \caption{SDC probability due to permanent faults in L1 instruction cache~\cite{10567770}.}
    \label{fig:hpca_l1i_permanent}
    %\vspace*{-5mm}
\end{figure}

Fig.~\ref{fig:hpca_l1i_permanent} illustrates the SDC probability outcomes for permanent faults in the L1 Instruction Cache across fifteen benchmarks of the MiBench~\cite{mibench_paper} suite for the three ISAs (Arm, x86, RISC-V). As shown in Fig.~\ref{fig:hpca_l1i_permanent}, the SDC probability ranges from 0.1\% to 2.3\% for Arm ISA, 0.1\% to 1.3\% for x86, and 0.3\% to 2.7\% for RISC-V ISA. These results are expected because a workload running with a persistent fault in any level of cache memory that stores instructions is unlikely to survive to the end and produce a corrupted output. Faults in most fields of instruction will primarily impact the execution flow or the instruction operands, and thus, lead to a crash~\cite{10151692}. On average across all benchmarks, the x86 ISA demonstrates the lowest SDC probability among the ISAs studied in this paper, while the RISC-V ISA shows the highest SDC probability in most benchmarks.

%\subsubsection{L1 Data Cache}

\begin{figure}[t]
    \centering
    %\vspace*{-1mm}
    \includegraphics[width=0.5\textwidth]{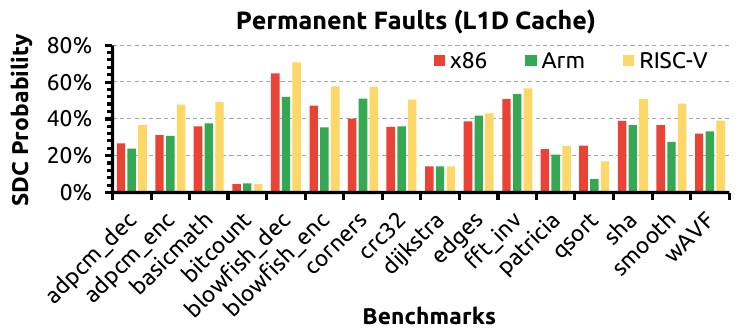}
    \vspace*{-4mm}
    \caption{SDC probability due to permanent faults in L1 data cache~\cite{10567770}.}
    \label{fig:hpca_l1d_permanent}
    %\vspace*{-5mm}
\end{figure}

Fig.~\ref{fig:hpca_l1d_permanent} displays the SDC probability results for permanent faults in the L1 Data Cache across the same fifteen MiBench benchmarks for the three ISAs (Arm, x86, RISC-V). As shown in Fig.~\ref{fig:hpca_l1d_permanent}, the SDC probability varies from 5.1\% to 53.3\% for Arm ISA, 4.4\% to 64.7\% for x86, and 4.4\% to 70.8\% for RISC-V ISA. On average across all benchmarks, the RISC-V ISA exhibits the highest SDC probability among all ISAs studied in this paper. It is important to note that the L1 Data Cache is considered unprotected in our experiments, i.e., there is no ECC-related protection scheme. The actual SDC probability can be much lower in real systems due to these protection mechanisms.

Overall, for the microarchitecture and workloads analyzed, the RISC-V ISA demonstrates a significantly higher probability of SDCs due to permanent faults compared to the other ISAs, i.e., Arm and x86. 

\section{Conclusions} %+ references 1 page
\label{sec:conclusions}

This paper provided an overview of AI-based security and RAS solutions in the computing continuum. The paper describes the comprehensive approach to malware
and hardware attack detection; as well as, the RAS features envisioned for future systems across the computing continuum. AI-based detection is shown to be a highly effective way to detect malware either statically or dynamically and with several ML methods. The paper also provides an analysis of the vulnerability to SDCs of L1 data and instruction caches for the same core but different ISA. The results show the importance of RAS features in future systems as the vulnerability to SDCs increases in each technology.

\bibliographystyle{IEEEtran}
\balance
\bibliography{bibliography}
\end{document}